 \documentclass[conference]{IEEEtran}

\ifCLASSINFOpdf
   \usepackage[pdftex]{graphicx}
\else
   \usepackage[dvips]{graphicx}
\fi
\usepackage{subfig}
\begin{document}

%
\title{Efficient Single Photon Absorption by Optimized Superconducting Nanowire Geometries}

\author{\IEEEauthorblockN{Mohsen K. Akhlaghi\\ and Jeff F. Young}
\IEEEauthorblockA{Department of Physics and Astronomy\\University of British Columbia\\
6224 Agricultural Road\\Vancouver, BC V6T1Z1\\Email: akhlaghi@phas.ubc.ca}
\and
\IEEEauthorblockN{Haig Atikian\\and Marko Loncar}
\IEEEauthorblockA{Harvard School of Engineering\\and Applied Sciences\\29 Oxford
Street\\ Cambridge, MA 02138 USA}
\and
\IEEEauthorblockN{A. Hamed Majedi}
\IEEEauthorblockA{ECE Department and\\Institute for Quantum Computing\\ University of Waterloo\\
200 University Ave West\\ Waterloo, ON Canada N2L3G1}
}


%


\maketitle

\begin{abstract}
We report on simulation results that shows optimum photon absorption by superconducting nanowires can happen at a fill-factor that is much less than 100\%. We also present experimental results on high performance of our superconducting nanowire single photon detectors realized using NbTiN on oxidized silicon. \footnote{Submitted to "Numerical Simulation of Optoelectronic Devices - NUSOD 2013" on 19-April-2013. \copyright 2013 IEEE. Personal use of this material is permitted. Permission 
from IEEE must be obtained for all other uses, in any current or future media, including reprinting/republishing this material for advertising or promotional purposes, creating new collective works, for resale or redistribution to servers or lists, or reuse of any copyrighted component of this work in other works.}
\end{abstract}



%
\IEEEpeerreviewmaketitle

\section{Introduction}
\begin{figure}[!b]
\centering
\includegraphics[width=2.2in]{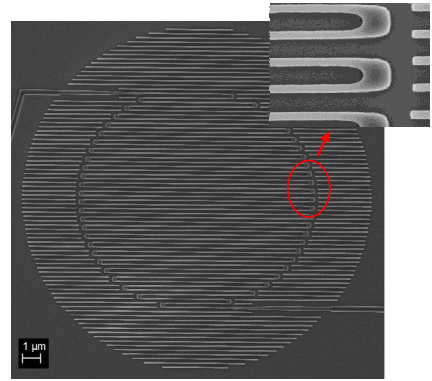}
\caption{Scanning electron microscope image of our detector presented in this paper. The diameter of the active area is 12.6$\mu$m.}
\label{fig:1}
\end{figure}

A superconducting nanowire kept well below its critical temperature, and biased close to its critical current is sensitive to single photons from visible to infrared. The nanowires are typically patterned from few nm thick superconducting film, deposited on a dielectric substrate. They form a meandering structure (see Fig.~\ref{fig:1}) that covers an area comparable with the mode field size of single mode fibers. This allows efficient coupling of photons to the detector active area, and thus in principle allows high detection efficiency.

It is intuitively believed that the higher the fill-factor (ratio of the nanowire width to the nanowire pitch), the higher the absorption of the photons. However, recent studies [1] on the effect of 180-degree turns on the detector performance suggests optimally designed bends (inset in Fig.~\ref{fig:1}) should be used in the detector structure; otherwise current would crowd at the bend positions, and excess dark counts would be generated. The most compact optimum 180-degree turn does not allow a fill-factor greater than 33\%, and this seems to mitigate the absorption of photons. Here, we present a detector design that gives maximum absorption at fill-factors considerably less than 100\%.

\section{Simulation and Design}
\begin{figure}[!t]
\centering
\includegraphics[width=2.3in]{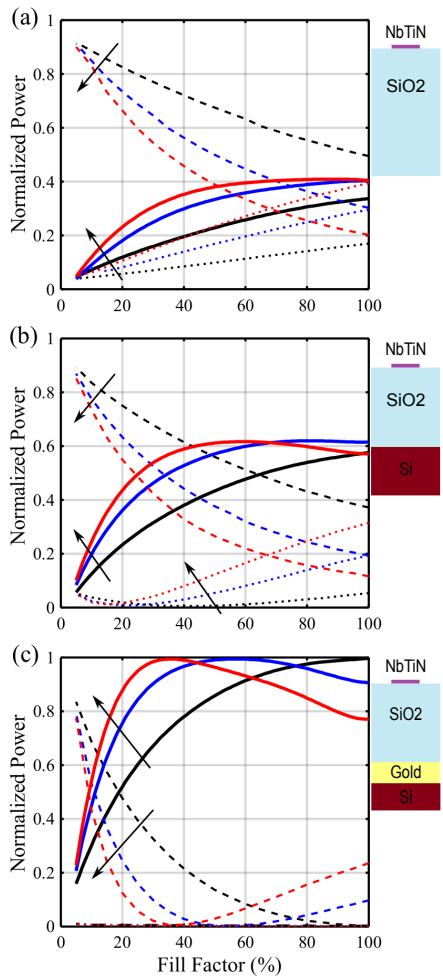}
\caption{Simulation results for the structures depicted in the insets. Absorption, Reflection and Transmission are shown by solid, dotted and dashed lines, respectively. The arrows show the direction of increase in the nanowire thickness: 4nm (black), 8nm (blue) and 12nm (red). Periodic boundary condition was used for the vertical sides.}
\label{fig:2}
\end{figure}

We use FDTD Solutions (from Lumerical Solutions Inc.) for the simulations. The first simulation is for periodic superconducting nanowires on top of semi-infinite SiO$_{2}$ substrate (see inset in Fig.~\ref{fig:2}(a)). NbTiN (real and imaginary parts of refractive index equal to 4.17 and 5.63 respectively) is assumed for the superconducting material. The nanowire width is fixed at 80nm, and the structure is simulated for 1550nm excitation with a polarization parallel to the nanowires. In Fig.~\ref{fig:2}(a), we plot simulated absorption (solid lines), reflection (dotted lines), and transmission (dashed lines) versus fill-factor for three different nanowire thicknesses. In all the cases, the absorption increases with fill-factor, while for higher thicknesses it approaches a plateau in which a decrease in transmission is counterbalanced with an increase in reflection.

The presence of the plateau by itself justifies designs with fill-factors considerably less than 100\%. That is because, once in the plateau, further increase of the fill-factor degrades the detector noise performance (as it increases the nanowires length) while it does not improve the absorption efficiency. However, for all the cases of Fig.~\ref{fig:2}(a), a 33\% fill factor (as required by the optimum 180-degree turns) is still outside the plateau.

At low fill-factors a large portion of the power is lost by transmission through the SiO$_{2}$ substrate. Therefore, it would be advantages to reflect back this power toward the nanowires. This brings us to our 2nd structure depicted in the inset of Fig.~\ref{fig:2}(b), where silicon, as a high index material, is used to reflect light. The Silicon substrate is assumed to be semi-infinite, and the thickness of the SiO$_{2}$ layer is set to $\lambda_{SiO_{2}}/4$, to let the lights reflected from Air-SiO$_{2}$ interface and the one reflected from the SiO$_{2}$-Si structure interfere destructively.  

Simulation results for overall absorption (solid lines), reflection (dotted lines), and transmission (dashed lines) are shown in Fig.~\ref{fig:2}(b). They show enhanced absorption compared to the previous structure. More significantly, the peak of absorption for thicker nanowires is at lower fill-factors, making them compatible with optimized 180-bends.  This happens because contrary to Fig.~\ref{fig:2}(a), the reflected power in the present case is not monotonically increasing with fill-factor: It shows a minimum at a fill factor much less than 100\%, where the two reflected deconstructing waves have the best balance.

Reflection by silicon improves absorption. However, considerable power is still lost through the substrate. Fig.~\ref{fig:2}(c) shows simulation results of the same structure but with a 60nm gold layer between the SiO$_{2}$ and the silicon substrate. Transmission goes to negligible values and absorption approaches 100\% for thicker nanowires at small fill-factors.

Although the above results show the possibility of rather ideal absorption at the desired fill-factors, this comes at expense of making the nanowires thicker. But the nanowires should be kept small to keep the single photon detection mechanism intact. Repeating the above simulations for other nanowire widths (40nm, 60nm and 80nm) we observe: the absorption at a fixed fill-factor is almost independent of the nanowire width. Therefore, it should be possible to increase thickness for having the highest absorption at the desired fill-factor, and then suppressed detection mechanism (due to increased nanowire cross section) would be restored by reducing the nanowire width while keeping fill-factor constant.

\section{Experimental Results}
\begin{figure}[!t]
\centering
\includegraphics[width=2.3in]{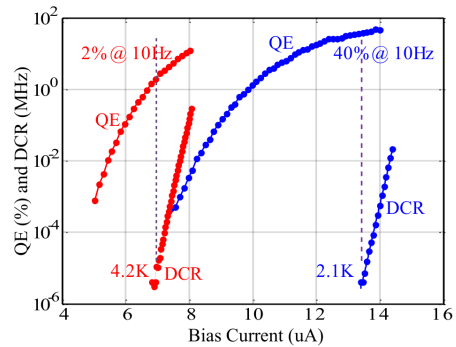}
\caption{Device Quantum Efficiency (QE) and Dark Count Rate (DCR) of our detectors measured at two different temperatures. Excitation wavelength is 1550nm and its polarization is parallel to the length of the nanowires. Measured critical temperature is 8.4K.}
\label{fig:3}
\end{figure}

Fig.~\ref{fig:1} shows a scanning electron microscope image of the detector we fabricated based on the design of Fig.~\ref{fig:2}(b). Nanowires are 80nm wide, 8nm thick NbTiN, and the fill-factor is 33\%. Installing the devices in an optical cryostat which allows precise calibration of the number of photons that hit the detector active area, we measured the noise and quantum efficiencies depicted in Fig.~\ref{fig:3}. At 2.1K, good performance is observed. We note the quantum efficiency approaches a flat level for higher bias currents. This justifies efficient operation of the detection mechanism in our rather thick but narrow nanowires. Therefore, the above design approach seems valid.

\section{Conclusion}
We showed the possibility of optimal absorption of photons by nanowires at low fill-factors, making them compatible with the optimized designs of 180-degree bends that do not allow fill-factors greater than 33\%. Experimental results confirms efficient and low noise single photon detection.


\section*{Acknowledgement}
We acknowledge the financial support of OCE, NSERC, IQC, and the Canadian Institute for Advanced Research. This work was performed in part at the Center for Nanoscale Systems
(CNS), a member of the National Nanotechnology Infrastructure
Network (NNIN), which is supported by the National Science
Foundation under NSF award no. ECS-0335765. CNS is part of Harvard
University.



%

\end{document}